# Group-kernel auto-calibration and group-patch k-space reconstruction: Fast MRI with time-variant $B_0$ kernels partitioned into time-invariant subsets

Rui Tian[1]*, Martin Uecker[2,3], Oliver Holder[1], Pavel Povolni[1], Theodor Steffen[1], Klaus Scheffler[1,4]

[1]High-Field MR center, Max Planck Institute for Biological Cybernetics, Tübingen, Germany [2]Institute of Biomedical Imaging, Graz University of Technology, Graz, Austria [3]BioTechMed-Graz, Graz, Austria [4]Department for Biomedical Magnetic Resonance, University of Tübingen, Tübingen, Germany

*corresponding author: rui.tian@tuebingen.mpg.de

**Abstract**

**Purpose**: Pushing MRI speed further demands more spatially-encoded information captured per unit time, e.g., by superimposing additional field modulations during oversampled readout. However, this can introduce calibration errors that degrade image quality. Thus, we propose a continuous field calibration approach. An efficient k-space reconstruction technique is also explored.

**Theory and Methods**: Our auto-calibration generalizes GRAPPA kernels to explicitly extract continuous $B_0$ modulation kernels, solving interpolation relationships between two ACS regions differing only in the extra field modulation. The k-space locations sharing the same instantaneous image-space modulation are grouped, so that subsets of time-invariant kernels can be separately estimated, as a generalized solution for Wave-CAIPI/FRONSAC-type scans. This view further inspires a k-space subregion-wise reconstruction technique, as an efficient alternative to conventional hybrid-space reconstruction. At 9.4T, FLASH accelerated by a local $B_0$ coil array and Wave-CAIPI were tested with retrospective undersampling.

**Results**: Artifact-free images were reconstructed, under diverse rapid $B_0$ modulation schemes, reaching maximum acceleration factors of 8-fold in 2D and 14.6-fold in 3D. Some nonlinear gradients modulation schemes reach similar sampling efficiency as linear gradients modulation. The proposed reconstruction shows potentials in further reducing reconstruction speed.

**Conclusion**: Rapid $B_0$ modulations and widely-adopted parallel imaging can share a common mathematical framework, and consequently, achieve similarly-robust reconstructions. Specifically, for scans using dynamic $B_0$ and static RF kernels, not only signal encoding, but also auto-calibration and reconstruction can be performed in k-space. This paves the way to robustly remove eddy currents, and explore more complex $B_0$ modulation strategies towards ultimate MRI speed.

## 1 Introduction

Despite decades of progress, MRI remains inherently slow due to the sequential nature of k-space sampling[1–3]. In a mathematical prospective of the forward model, there are two strategies to fundamentally speed up MRI acquisitions: physically collecting more spatially-encoded information within shorter scan times[4–14], or solving the forward model with advanced reconstruction algorithms to suppress undersampling artifacts[15–19]. Within the broad spectrum of fast MRI strategies, this paper introduces a novel field calibration technique termed group-kernel auto-calibration, and an efficient reconstruction method termed group-patch k-space reconstruction, to facilitate a critical yet currently underutilized image acceleration approach based on rapid $B_0$ field modulations. These modulations can be linear gradient modulations as in Wave-CAIPI, second- and third-order nonlinear gradient modulations as in FRONSAC, or more general and flexible modulation patterns produced by local $B_0$ coil array.

For example, in Cartesian MRI, the readout Nyquist sampling rate is determined by the FOV and the gradient strength for frequency-encoding (i.e., k-space traversal speed). When the scanner's ADC can sample faster than this Nyquist-limit, the readout oversampling does not add new and useful spatial information, and is usually removed by averaging adjacent oversampled data. However, this otherwise redundant oversampling can be exploited by superimposing additional rapid magnetic field modulations onto the frequency-encoding gradient, thereby capturing extra phase-encoded information within the same readout duration. Specifically, the k-space line-by-line trajectory can be perturbed into 2D zig-zag (e.g., Bunched-phased-encoding/BPE) patterns[10,11], 3D corkscrew patterns (e.g., Wave-CAIPI[12]), or spread into time-varying k-space "stamps" (e.g., FRONSAC[20–22], local $B_0$ coil modulations[23,24], temporally-varying RF receiver profiles[25,26]), thereby boosting information influx to the signal encoding model. According to the convolution theorem, each k-space sample at a given time can be viewed as a weighted sum over the underlying k-space values in a local neighborhood. In this paper, we refer the corresponding set of weights induced by rapid $B_0$ fields as the $B_0$ modulation kernel, which is the Fourier transform of the image-space phase modulation pattern at that time. This $B_0$ kernel provides a unified description of the effective $B_0$ gradient modulations in k-space at a time, ranging from Dirac delta function in shifts to distributed "stamp".

By comparison, it is common to speed up acquisition by accelerating "long-range" k-space traversal using high-performance gradients, whereas local "wiggling" the trajectory using additional continuous $B_0$ gradient modulations remains much less investigated, given a large yet underexplored space of nonlinear gradient modulation fields[27,28]. Unfortunately, these time- and space-varying modulations require accurate calibration for robust image reconstruction, similar to image-space RF receiver sensitivity maps or k-space GRAPPA kernels (i.e., kernels of RF receive fields) that vary across receiver channels in parallel imaging. It can be even more challenging, because continuous $B_0$ field modulations can introduce additional complications, such as substantial trajectory deviations due to complex spatial field patterns, time delay and eddy currents, as well as mechanical vibrations – that are usually absent in the time-invariant RF receivers. As a result, this image acceleration strategy could be hindered for broad adoption.

To estimate the additionally-imposed spin phase modulations, standard $B_0$ field mapping scans for each linear/nonlinear gradient element (e.g., with blip or DC current)[29,30,24] can be performed, combined with current monitoring and kernel-based extrapolation[24]. But these maps cannot capture dynamic effects (e.g., the eddy currents during $B_0$ modulations), possibly ending up with residue fold-over artifacts. Alternatively, a CSI-type sequence can measure a more realistic spatiotemporal model along readout by phase-encoding all spatial pixels[30–32]. However, it is time-consuming, thermally and mechanically stressful for hardware, requires careful sequence debugging and coil cooling to match calibration-to-imaging conditions, and remains vulnerable to noise and measurement errors. For special case of linear gradient modulations (e.g., Wave-CAIPI), field calibration can be simplified using single-slice projection data[12], with optimization of a few parameters ("partial" auto-calibration)[33] in reconstruction. Nevertheless, robust, widely-deployable field-modulated scans should prefer a complete auto-calibration strategy to estimate a full spatiotemporal model directly from imaging data.

Therefore, we propose group-kernel auto-calibration to generalize self-estimation of time-invariant RF receive ($B_1^-$) kernels to continuous $B_0$ modulation kernels, auto-calibrated in separate groups[34]. Inspired by this view of $B_0$ kernel groups, we further propose group-patch reconstruction to reconstruct field-modulated data more efficiently via interpolations of undersampled k-space data in patches. Essentially, both are built on a novel perspective that, the full set of time-variant k-space $B_0$ kernels can be partitioned into time-invariant subsets across selected k-space locations.

For calibration, from a standard auto-calibration-signal (ACS) region and a field-modulated ACS region, a series of time-invariant $B_0$ modulation kernels can be separately estimated, by grouping data at all k-space locations that share identical interpolation relationship between the two ACS regions. These kernels together yield the full set of time-varying $B_0$ kernels, providing a complete spatiotemporal phase evolution model that captures the cumulative effects from field modulations and all undesirable imperfections. This also represents a k-space counterpart to a recent image-space auto-calibration technique that estimates the pixelwise point-spread-function (PSF) for Wave-CAIPI/FRONSAC[22]. In comparison, when the nonlinear modulations vary slowly across space, k-space kernel extraction benefits from averaging across a large number of linear equations (i.e., due to kernel shifts), and requires only narrow k-space bands rather than full-resolution images as the auto-calibration regions, elegantly inheriting GRAPPA's robustness. Part of our auto-calibration concepts appears similar to a recent nearly-continuous extension[35] of dual-polarity GRAPPA[36] for EPI, together delineating a more complete picture for linear and nonlinear gradient modulations.

For the proposed group-patch reconstruction, undersampled k-space data can be further grouped into patches, each sampled by a redefined subset of time-varying $B_0$ kernels. Building on a recent MRI sampling theory describing generalized k-space interpolations based on reproducing kernel Hilbert space (RKHS)[37], we compute, for each kernel subset, a single interpolation matrix that can map any undersampled k-space patch in that group to a fully-recovered patch. This interpolation matrix is then efficiently reused to reconstruct all undersampled patches, which are concatenated into a full k-space.

To experimentally test our approaches, 2D and 3D FLASH scans are accelerated by an 8-channel local $B_0$ coil array as a relatively-arbitrary field-generating hardware, as well as the scanner's linear gradients (i.e., for Wave-CAIPI). Without dense array of field probes[38] – which cannot fit between the local $B_0$ array and the RF coil – our k-space auto-calibration method substantially improves reconstruction quality. It remains robust across linear and nonlinear modulation schemes, along all spatial orientations in a 9.4T human MRI scanner, even under occasional hardware noise. Furthermore, the k-space patch-wise reconstruction speeds up reconstruction compared to conventional Wave-CAIPI type hybrid-space reconstruction. Overall, our methods enable fast yet robust MRI with rapid $B_0$ field modulations, while its fundamental acceleration limits depend on the time-integral of $B_0$ spatial gradients, rather than the RF ($B_1$) wavelengths[39].

## 2 Theory

### 2.1 System overview at 9.4T

In this study, 2D and 3D FLASH line-by-line scans are accelerated by superimposing additional rapid $B_0$ field modulations onto the frequency-encoding gradient during signal readout, using either a custom-built 8-channel local $B_0$ coil array placed outside the RF coil, or scanner's linear gradient as Wave-CAIPI. In Figure 1, sinusoidal currents are applied to each channel of local $B_0$ coils or scanner gradients, imposing either local or global spin phase modulations, improving pixel disentanglement in image reconstruction using retrospectively undersampled k-space data. In this paper, the rapid $B_0$ modulations are identical across phase-encoded steps (i.e., TRs) for simplicity.

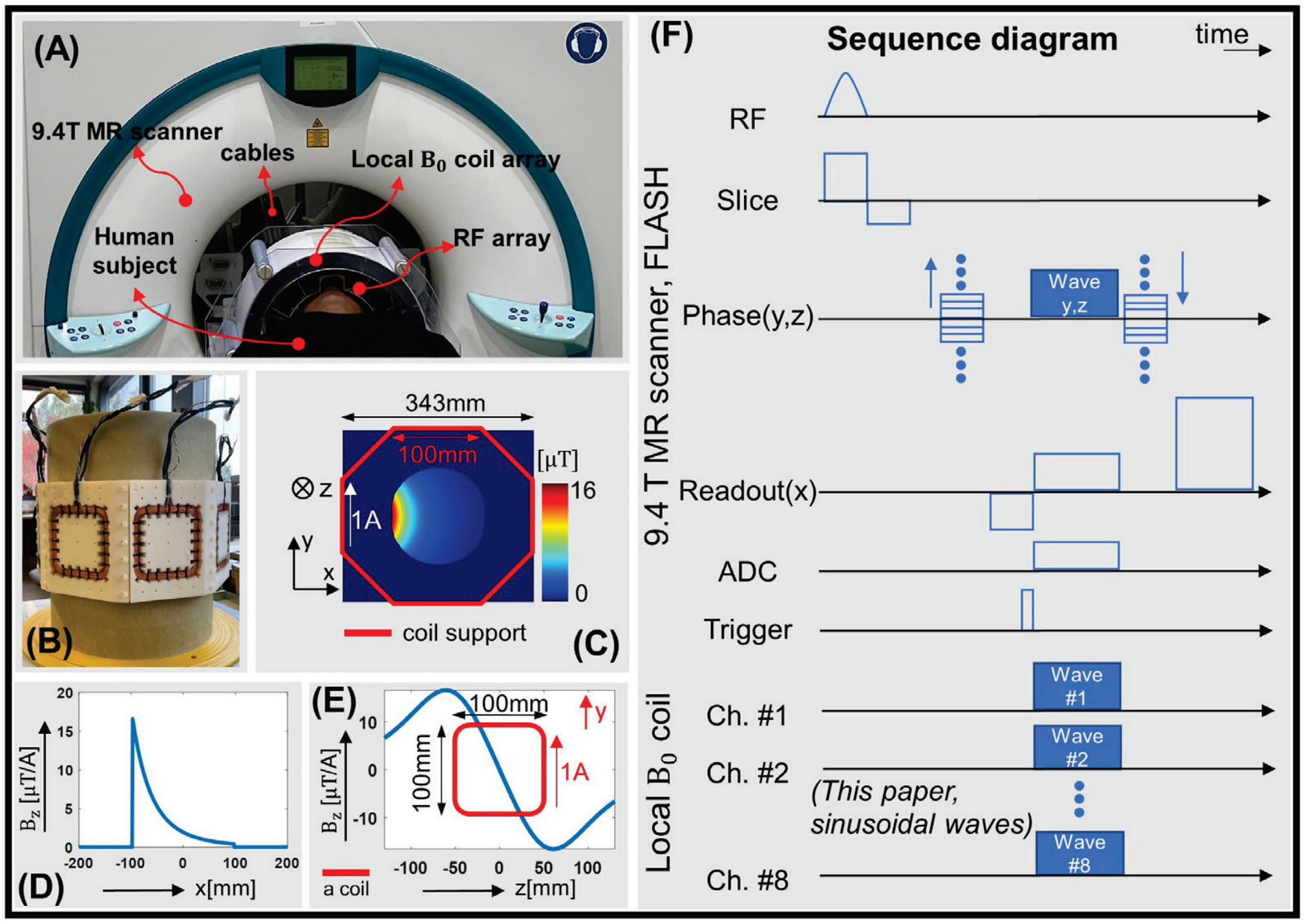

Figure 1. System illustration, see more details in the reference[24]. (A) The hardware setup in the scanner. (B) The local $B_0$ coil array under construction (a replica to the reference). (C) Simulated magnetic field map given 1A DC current in a single local $B_0$ coil, which generate up to around 16 μT within the effective field of view (FOV) defined by the inner boundary (around 97mm to the FOV center) of a RF array. (D) The 1D plot along the central horizontal line in (C). (E) The variation of the in-plane maximal field strength generated by 1A in a local coil along z direction, with the relative position to the coil winding indicated in red. (F) Pulse sequence diagram for accelerated MRI scans with rapid $B_0$ field modulations. In a standard FLASH sequence, during signal readout, either the local $B_0$ coil array or scanner's phase-encoding gradients are switched on with current waveforms. Theoretically, the waveforms in the 8 local $B_0$ channels can be designed independently. Here, we apply sinusoidal currents for simplicity, to induce additional localized spin phase modulation on the objects to accelerate k-space acquisitions.

## 2.2 k-space $B_0$ kernels to model linear and nonlinear gradients modulations

In conventional signal model[8,40], each k-space sample $S_n(t)$ acquired at time t from RF receiver channel $n$ is obtained by first multiplying the spin density map by the RF receiver sensitivity map, the phase map induced by rapid $B_0$ field modulations, the phase map induced by linear gradient for frequency encoding, and then summing all the image-space contributions into a single k-space value at that time.

$$S_n(t) = \int c_n(r)\rho(r)\, exp\left\{-i\left\{\gamma \sum_{\lambda}\left[B_\lambda(r)\int_0^t I_\lambda(\tau)\, d\tau\right] + k(t)r\right\}\right\} dr, (1)$$

with $k(t) = \gamma \int_0^t g(\tau)\, d\tau$ as the k-space trajectory term (i.e., in rad/m)

where $\gamma$ is the gyromagnetic ratio (i.e., in rad/s/T), r is the spatial location, $c_n(r)$ is the $n^{th}$ RF receiver sensitivity map, $g(\tau)$ is the shaped pulse of the linear gradients, $B_\lambda(r)$ is the $B_0$ field distribution produced by unit current in the $\lambda^{th}$ coil that generates rapid field modulations (local $B_0$ coil or phase-encoded linear gradient), $I_\lambda(\tau)$ is the programmable current waveform in the $\lambda^{th}$ modulation coil. The spatial distribution of rapid $B_0$ fields at a time can be linear or nonlinear.

Multiplying by the linear gradient induced phase map and summing over all image-space positions is equivalent to evaluating the Fourier transform at a k-space location corresponding to time $t_0$. According to the convolution theorem, the k-space signal can thus be expressed as the convolution of the Fourier transforms of the three maps, evaluated at that k-space location $\mathrm{k}(\mathrm{t}_0)$[19]:

$$S_n(t_0) = \left\{ \underbrace{\mathcal{F}\{c_n(r)\}}_{RF\ receive\ kernel} * \underbrace{\mathcal{F}\{\rho(r)\}}_{original\ k-space} * \underbrace{\mathcal{F}\left\{ exp\left\{ -i\gamma \sum_{\lambda} \left[ B_\lambda(r) \int_0^{t_0} I_\lambda(\tau)\, d\tau \right] \right\} \right\}}_{B_0\ modulation\ kernel} \right\}(t_0), (2)$$

where asterisk $*$ is the convolution operation. In this representation, the original k-space object is convolved with a RF receive kernel, so that each k-space sample in channel $n$ becomes a weighted sum of a local neighborhood of the original k-space object. A subsequent convolution with the $B_0$ modulation kernel further mixes neighboring samples in receiver channel $i$, so that the final value is a weighted sum of another local neighborhood of the channel-specific k-space data. Note that, the term “GRAPPA kernel”[9] usually refers to the k-space kernels used to interpolate missing samples from undersampled data in multiple receivers. Here, suggested by the RKHS MRI sampling theory[24,37], we use “RF receive kernel” as well as “$B_0$ kernel” in a more general sense to denote any k-space interpolation kernel that maps between the original and the acquired k-space in either direction, due to RF receiver or $B_0$ modulation fields. Across different k-space locations, the RF kernel is shift-invariant and the $B_0$ kernel is shift-variant, whereas across distinct RF receiver channels, the RF kernel is shift-variant and the $B_0$ kernel is shift-invariant. Thus, when Equation (2) is evaluated at different time points and for different receiver channels, the corresponding kernel weights might differ.

### 2.3 Continuous field calibration using kernels in time-invariant groups

The proposed group-kernel auto-calibration technique is based on interpolating k-space data using kernels – which is also a central concept for RF receivers calibration in GRAPPA and ESPIRiT[41]. In the original GRAPPA formulation, k-space data are reconstructed by multiplying acquired k-space data neighborhoods with receiver-specific GRAPPA kernels and summing to interpolate missed k-space data. These kernels are estimated from a fully-sampled ACS region. Alternatively, as in Figure 2A, fully-sampled RF receiver kernels with full k-space support (the black, or the red lines) can be estimated by learning interpolation relationships between local neighborhoods of one receiver (the gray, or the orange block) and corresponding data points from another receiver (used as a reference, the blue block with black outline). These fully-sampled RF kernel coefficients can be Fourier transformed into image-space receiver sensitivity maps, enabling SENSE-like reconstructions similar to ESPIRiT. To estimate the RF kernel weights, a large number of overdetermined linear equations (or training datasets for neural networks) are generated by assuming the kernels are shift-invariant, while the data themselves remain shift-variant across k-space.

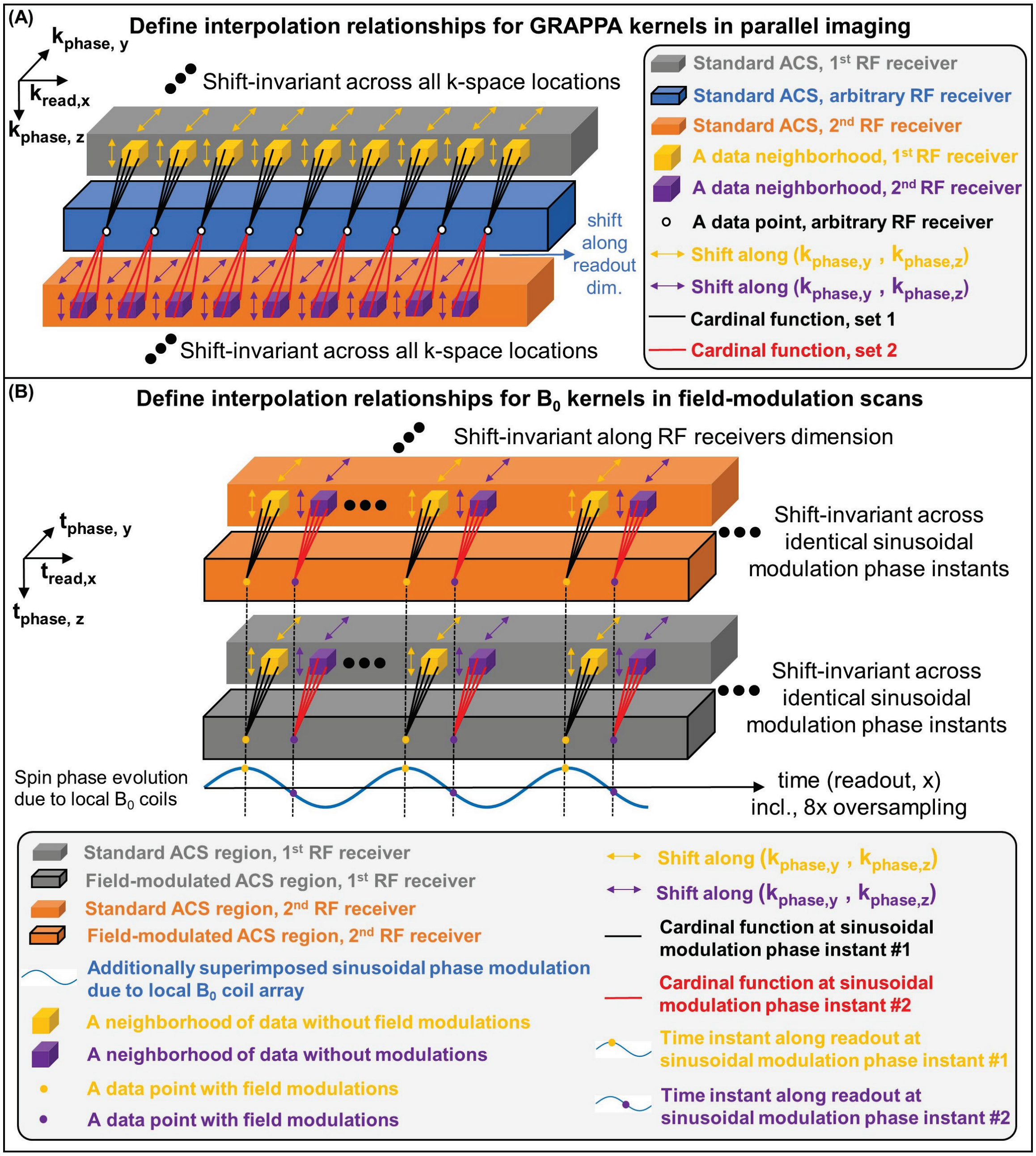

Figure 2. Auto-calibration from parallel imaging to continuous modulations of $B_0$ fields, both by extracting k-space kernels (RF or $B_0$) as the Fourier transform of image-space signal modulations. (A). Review of the interpolation relationships governing auto-calibration of GRAPPA weights, interpolating a neighborhood of k-space data encoded by distinct RF receivers to a data point from a RF receiver. A set of GRAPPA weights is shift-invariant along three k-space axes (i.e., $k_x$, $k_y$, $k_z$), unique to specific pair of RF receivers. (B) Auto-calibration of rapid $B_0$ modulations using e.g., sinusoidal modulations of our 8-channel local $B_0$ coil array or scanner's phase-encoding gradients. A $B_0$ modulation kernel at a particular phase of the sinusoidal oscillation cycle is shift-invariant along two phase encoding axes (i.e., $t_y$, $t_z$), the RF receivers' dimension, and among different readout time instants sharing the same phase of the sinusoidal modulation cycle.

Similarly, as in Figure 2B, the rapidly varying $B_0$ modulation fields lead to shift-variant $B_0$ kernels across k-space along the readout direction in Cartesian line-by-line scans, while being shift-invariant across receiver channels. To estimate such time-varying kernels (black and red lines), the proposed group-kernel auto-calibration uses two ACS blocks with readout oversampling: a standard ACS (orange and gray blocks without black outline), and a field-modulated ACS (orange and gray blocks with black outline). Both datasets can also serve as imaging echoes during reconstruction. We partition k-space locations into groups sharing identical spin modulation states, readout timepoints with the same sinusoidal modulation phase across all phase-encoding lines and RF receivers. Within each group, many source-target data pairs $f_i^{i,\dot{t}}(t)$ and $g_i(t)$, supported on neighborhoods $\dot{t} \in D_{\dot{t}}$, are interpolated using the same kernel $w_i^{i,\dot{t}}(t)$, regardless of different k-space locations t as the neighborhood centers. This defines a subset of interpolation relationships.

$$g_i(t) = \sum_{\dot{t} \in D_{\dot{t}}, i} f_i^{i,\dot{t}}(t) w_i^{i,\dot{t}}(t)\,, (3)$$

A small sliding window moves across the ACS data subset in this group to assemble a linear system for kernel estimation:

$$F^H \widehat{U} = G^H, (4)$$

$$\text{or, } M_{ACS} \widehat{U} = R_{ACS}, (5)$$

$$\text{with} \begin{cases} M_{ACS} = FF^H, \ \in \mathbb{C}^{D\times D}, (6) \\ R_{ACS} = FG^H, \ \in \mathbb{C}^{D\times 1}, (7) \end{cases}$$

The matrix $F \in \mathbb{C}^{D\times S}$ consists of S neighborhood vectors (each has D samples) from the standard ACS, while $G \in \mathbb{C}^{1\times S}$ contains the corresponding target samples from the field-modulated ACS. The kernel $\widehat{U} \in \mathbb{C}^{D\times 1}$ can be solved using pseudoinverse, truncated-SVD (tSVD), LSQR, or utilizing neural network. Superscript $(\cdot)^H$ denotes conjugate transpose. Optionally, left-multiplying the Equation (4) by $F$ yields the normal equations in Equation (5) with $M_{ACS} \in \mathbb{C}^{D\times D}$ and $R_{ACS} \in \mathbb{C}^{D\times 1}$.

Repeating this procedure across different kernel groups (i.e., different $w_i^{i,\dot{t}}(t)$) yields the full set of time-varying $B_0$ kernels, which after Fourier transform become spatiotemporal spin-phase-evolution maps. These maps capture the cumulative effects of rapid field modulations, eddy currents, gradient delays, concomitant fields, and subtle motions. The standard ACS can be acquired first, so that inversion of Equation (4) or (5) can start earlier. Optionally, receiver channels can be combined before group-kernel auto-calibration. Since kernel calibration per group is independent, the computation is parallelizable.

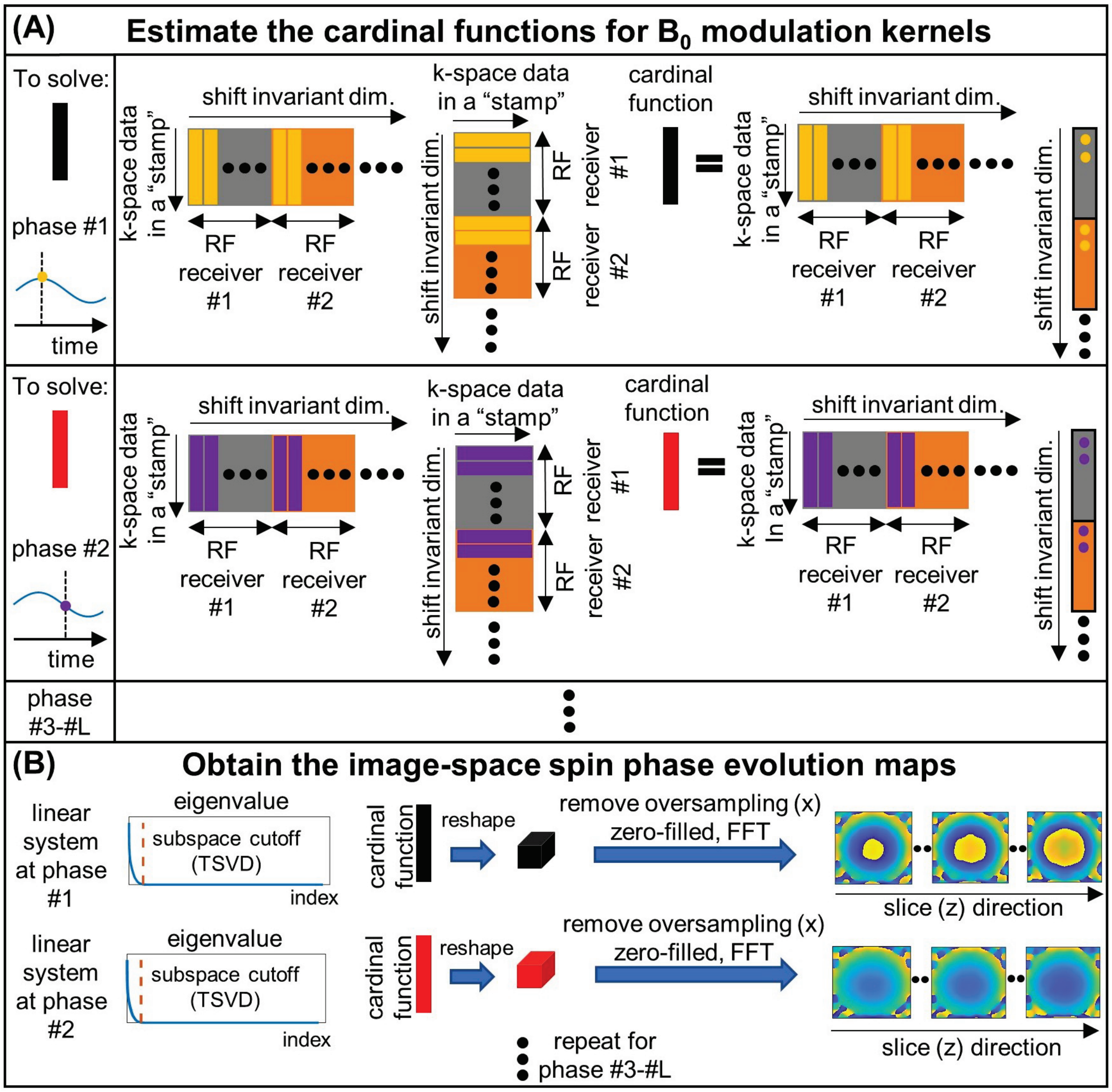

Figure 3. Grouped-kernel auto-calibration based on k-space subregions that span multiple kernel groups. Some symbols are defined in Figure 2. (A) Step 1: Define the kernel interpolation relationships, e.g., a set of linear equations, to interpolate source neighborhoods from the standard ACS to target points from the field-modulated ACS. This interpolation direction can also be reversed. (B) Step 2: Solve the cardinal functions at each phase of a sinusoidal oscillation cycle using a numerical solver, which theoretically, can be linear, nonlinear, or based on neural networks. For example, optional eigenspace filtering can be applied, e.g., using truncated singular value decomposition (TSVD). The interpolation weights are then reshaped, zero-padded and Fourier transformed to yield spin phase evolution maps.

The same principle can be applied in reverse – interpolating from the field-modulated ACS to the standard ACS, similar to classical GRAPPA calibration. This pathway may not easily yield a full spatiotemporal encoding model in a straightforward manner. If extended to EPI as an extreme case of Wave-CAIPI (i.e., a large k-space "sweep", one or a few shots), it appears similar to a nearly-continuous extension of dual-polarity GRAPPA[35], where two (equivalent) field-modulated ACS blocks with opposite EPI readout polarities that can jointly estimate $B_0$ and RF receiver kernels for k-space reconstruction.

### 2.4 Interpolating k-space patches spanning time-invariant modulation segments

The proposed group-patch reconstruction (i.e., image reconstruction by interpolating k-space patches in groups) is based on cardinal functions from the RKHS formalism of MRI. These cardinal functions act as generalized k-space interpolation weights, mapping a patch of undersampled k-space data (source) to another patch of fully-sampled data. In its original form, computing the full k-space cardinal function from the entire signal model of Equation (1) can be computationally prohibitive.

However, previous work has shown that, under nonlinear gradient modulation that is periodic along the readout (and often identical across phase-encoded steps), the cardinal function can be evaluated within a small k-space region, while still reflecting the encoding efficiency of the full k-space. Inspired by this, we partition the undersampled k-space into patches of fixed size, each covering an integer number of modulation periods and sampled by the same subset of $B_0$ kernels. Consequently, the cardinal function matrix needs to be computed only once for such a kernel subset, since all patches are sampled by that kernel subset and share identical interpolation weights to reconstruct fully-sampled k-space. This reduces the problem to the inversion of a small linear system for a localized k-space subregion, and eliminates repetitive computation at different k-space locations where the underlying field modulation segments are identical across patches.

First, for a pair of source and target patches, we construct two encoding matrices with acquisition indices $(t, i)$ and $(\bar{t}, j)$. Each row represents a spatial encoding map at a time instant, flattened into a 1D vector. The source encoding matrix $E_{\bar{t},j}$ contains composite spatial encoding maps corresponding to the sampling of the source patch. The target encoding matrix $E_{t,k}$ is a truncated DFT matrix that represents Fourier sampling of the target patch, optionally weighted by a receiver channel sensitivity map $c_k(r)$.

$$E_{\bar{t},j} = c_j(r)exp\left\{-i\left\{\gamma\sum_{\lambda}\left[B_\lambda(r)\int_0^{\bar{t}} I_\lambda(\tau)\,d\tau\right] + k(\bar{t})r\right\}\right\}, (8)$$

$$E_{\dot{t},k} = c_k(r)exp\{-ik(\dot{t})r\}, (9)$$

If reconstructing within a single RF receiver, we set $k = j$. If reconstructing data jointly encoded by RF coils and rapid $B_0$ modulations, we set $c_k(r) = 1$ to interpolate onto a reference Fourier grid without RF coil weighting.

Second, we compute the Gram ($\mathrm{M}_{sp}$) and cross-Gram ($\mathrm{R}_{tp}$) matrices from these two encoding matrices, and loaded into the RKHS formalism to obtain the cardinal function matrix U, i.e., the interpolation weight matrix that maps the source patch onto the target patch.

$$\begin{cases} M_{sp} = E_{\bar{t},j}E_{t,i}^*, \in \mathbb{C}^{TN\times TN}, (10) \\ R_{tp} = E_{\bar{t},j}E_{\dot{t},k}^*, \in \mathbb{C}^{TN\times \dot{T}}, (11) \end{cases}$$

Specifically, we solve:

$$M_{sp}U = R_{tp}, (12)$$

where $U \in \mathbb{C}^{D\times\widehat{D}}$ , $D$ and $\widehat{D}$ are the sample numbers in the source and target patches respectively. The power function representing k-space approximation errors can also be derived from the estimated cardinal function:

$$P^{\dot{t},k} = \sqrt{\left(E_{\dot{t},k}E_{\dot{t},k}^* - U^*E_{t,i}E_{\dot{t},k}^*\right)_{diag}}, (13)$$

Third, the calculated cardinal function matrix is used to for reconstructing a target patch, by matrix-vector product.

$$\hat{F}_{tp}^{H} = F_{sp}^{H}U, (14)$$

where $\hat{F}_{tp} \in \mathbb{C}^{\hat{D}\times 1}$ is the vectorized reconstructed k-space patch, $F_{sp} \in \mathbb{C}^{D\times 1}$ is the vectorized undersampled k-space patch. In practice, the source patch $(\cdot)_{sp}$, is chosen slightly larger than the target subregion $(\cdot)_{tp}$ to avoid boundary errors.

Since all patches span time-invariant $B_0$ modulation segments and share identical interpolation weights to fully-sampled k-space, reconstruction can proceed in parallel by interpolating each undersampled source patch to its corresponding fully-sampled target patch. The reconstructed patches are assembled to form the full k-space data, which is Fourier transformed to yield the final reconstructed image.

Unlike GRAPPA/SPIRiT reconstructions, where interpolation weights are tied to the local training kernels, the RKHS framework allows direct interpolation onto Fourier grid regions with arbitrarily selected k-space ranges. This enables flexible, region-by-region reconstruction decoupled from the learned kernels. Compared to global RKHS computation, our approach allows interpolation to k-space subregions tailored for modulation periods of rapid $B_0$ field, substantially reduces the computational cost. Similar subregion interpolation strategies have been explored for non-Cartesian gridding, but with different implementations.

Compared with the conventional hybrid-space reconstruction (originally developed for Wave-CAIPI) that decouples only the phase-encoding dimensions, our patch-wise reconstruction can decouple all k-space dimensions, including readout with time-varying $B_0$ field modulations. This can flexibly slice reconstruction data into smaller chunks, and substantially reducing computational burden further.

Theoretically, if applied to EPI, the same principle supports continuous trajectory calibration that can be decoupled from parallel-imaging reconstruction. This avoids embedding the full signal model into SENSE-type reconstructions, which would otherwise become computationally expensive.

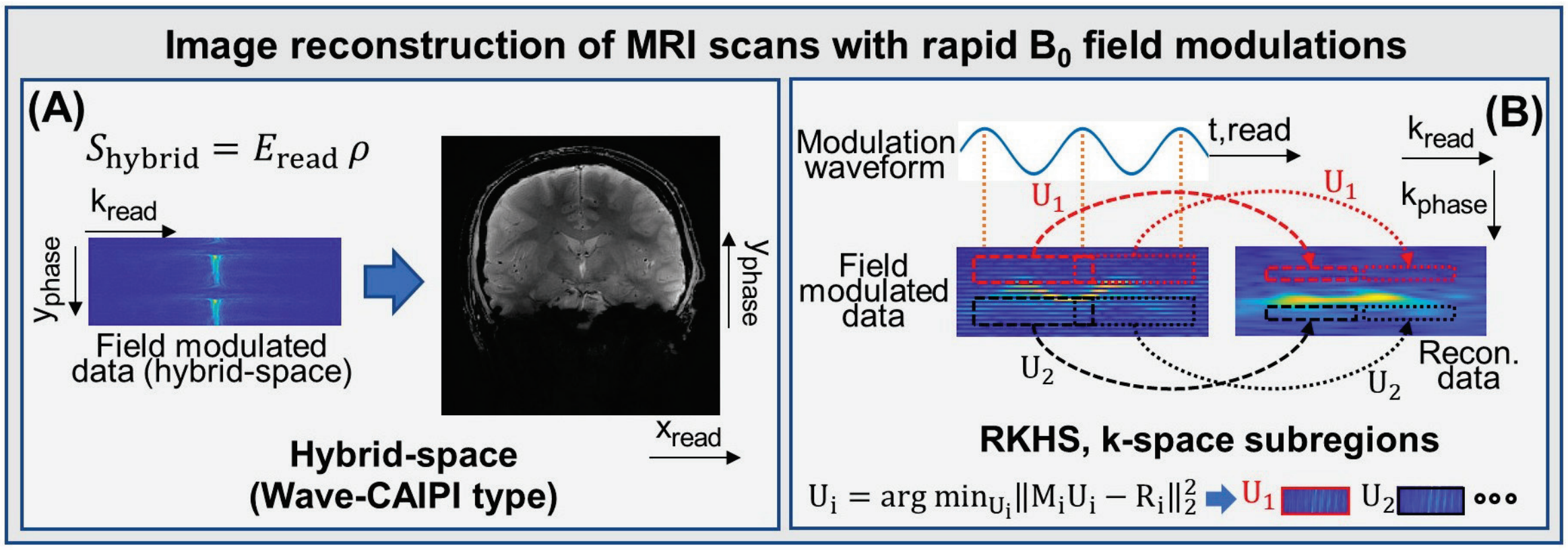

Figure 4. (A) In hybrid-space reconstruction, the phase-encoding dimension(s) are solved by FFT, while the readout forward model is solved by general inversion. Note, the k-space dataset and the reconstructed image only serves as illustration purposes, and size-ratio between their figures is not realistic. (B) In group-patch reconstruction, the k-space data are decoupled in subregions, spanning periodically occurring $B_0$ kernels, along both phase-encoding and readout dimensions. Cardinal functions for distinct subregions (index i) are calculated based on Equation (10-12), and applied for k-space interpolation based on Equation (14). Appropriately defined subregion can further reduce peak memory and reconstruction time.

## 3 Methods

Ex-vivo and in-vivo FLASH scans, with and without $B_0$ gradient modulations, were performed on a 9.4T whole-body human scanner (Siemens Healthineers, Germany), equipped with SC72 gradients with maximum slew-rate and absolute-strength of 200T/m/s and 70mT/m. To compare various rapid $B_0$ modulation schemes, an 8-channel local $B_0$ coil array (24.1kg before epoxy fixation; 33.5kg after, eight 10cm x 10cm square loops, 14 turns each, approximately 41 μH measured at 7kHz) was placed outside of a 16-transceive/32-receive RF coil array (modified from the reference[42], CP-mode transmission, joint acceleration), fixed to the patient table. Several healthy adult volunteers were scanned under institutional ethics approval and informed consent.

During FLASH frequency-encoding, scanner's ADC was 8x oversampled with dwell-time of $3\mu s$. In $B_0$ field-modulated scans, either the local $B_0$ coils or the scanner's phase-encoding (linear) gradients were driven with sinusoidal currents, implemented with a customized IDEA sequence synchronized with external power-amplifier control, and a Pulseq[43] sequence, respectively. For each experiment, fully-sampled k-space dataset with and without (reference) field modulations were acquired. Standard ACS regions, field-modulated ACS regions and undersampled k-space data were then cropped retrospectively, which allowed systematic (different undersampling factors) reconstruction tests with a worst-case guarantee due to possibly stronger subject motions.

For ex-vivo phantom scans in Figure 5-7, multi-slice 2D FLASH in 1.06mm$^2$ were scanned with FOV 190x190mm$^2$, matrix-size 180x180, slice-thickness 3mm, readout-bandwidth 230Hz/Pixel, readout - duration 4.32ms, TE/TR 4-5ms/30ms. The $B_0$ coils currents were 7.41kHz/40$A_{pk}$. In Figure 8, in-vivo 3D FLASH in 1.7x1.7x1.5mm$^3$ was performed with FOV 220x220x132mm$^3$, matrix-size 128x128x88 (slice oversampling 37.5%), readout-bandwidth 330Hz/Pixel, readout-duration 3.072ms, TE/TR 4ms/10ms. The $B_0$ coils currents were 9.80kHz/37$A_{pk}$. In Figure 9, in-vivo 2D FLASH was scanned with FOV 250x250mm$^2$, matrix-size 280x280, slice-thickness 3mm, readout-bandwidth 150Hz/Pixel, readout-duration 6.720ms, TE/TR 6.07ms/30ms. The $B_0$ coils currents were 7.41kHz/36$A_{pk}$. In Figure 10 (using local $B_0$ array), in-vivo 3D FLASH was scanned with FOV 230x230x88mm$^3$, matrix-size 230x230x112 (slice-oversampling 1.27), readout-bandwidth 180Hz/Pixel, readout-duration 5.520ms, TE/TR 6ms/30ms. The $B_0$ coils currents were 7.41kHz/36$A_{pk}$. In Figure 10 (wave-CAIPI), in-vivo 3D wave-CAIPI implemented in Pulseq was designed similarly as the 3D FLASH with $B_0$ coil array (i.e., identical FOV, matrix-size, readout-bandwidth, readout-duration, TE/TR. Nevertheless, the slew-rates and strengths of the oscillating phase-encoding gradients (y-z) were (3.4-5.3) mT/m, 4.17kHz for non-overlap wave-trajectory with 24 wave-cycles, and (5.3-8.2) mT/m, 2.78kHz for overlapped wave-trajectory with 16 wave-cycles.

The kernel-size ranged from by 6 to 19 along one dimension, with typically 15-60 kernel shifts along each of readout, phase-encoding and RF receivers dimensions, depending on ACS sizes. Removing readout oversampling in the selected source neighborhood before solving interpolation relationships is much more efficient. Typical calibration times are shown in the Figure 8-10.

Most images were still reconstructed in hybrid-space (i.e., FFT resolving phase-encoding dimensions, forward model resolving readout dimension) similar to wave-CAIPI, with optional joint data compression (JCS.) based on $B_0$-RF kernels to accelerate compressed-sensing reconstruction. The joint compression will be described in a separate paper. In Figure 11(C), the 2D image was obtained via group-patch reconstruction with k-space subregions interpolations, without compression and parallel imaging, as proof-of-concept implementations.

The linear systems for kernel estimations and least-square ($L_2$) reconstructions were solved by LSQR, and the CS-reconstructions were solved by FISTA with $L_1$ regularizations based on total-variation and wavelet. The variable-density undersampling mask was generated by BART[44]. The difference image is calculated based on absolute value of the magnitude image difference, normalized by the range of the reference magnitude image. The normalized-root-mean-square-error (NRMSE) is normalized by the range of the reference magnitude image. The computation was performed offline in MATLAB using HPC-servers (CPU, 2.35GHz, 64 cores), with 1TB RAM option for 3D scans.

## 4 Results

### 4.1 Robust field calibration across different spatial axes

In Figure 5, the proposed grouped-kernel field calibration technique is compared with the previous calibration approach[24] that combines individual $B_0$ mapping for each local $B_0$ coil, current monitoring of sinusoidal waveforms, and ESPIRiT-based extrapolation to high-resolution grid. To evaluate calibration robustness in the presence of potentially complex spatiotemporal eddy current patterns, 2D field-modulated FLASH scans were performed with different slice orientations, based on fully-sampled k-space and least-square reconstruction – ensuring that any image artifacts result from field calibration rather than undersampling. Note, the RF shield between our RF and local $B_0$ coils was not optimized for cutting eddy currents flowing along z-axis.

In the transverse plane, when the local $B_0$ coils generated a nearly-linear gradient oscillation along the phase-encoding dimension (i.e., bunched-phase-encoding) with the linear gradient (“LG”) current configuration (with 0-deg and 180-deg sinusoidal phase offsets), both calibration methods produce artifact-free reconstruction, as in previous study. However, under more complex nonlinear field modulations (e.g., the current configuration “octupolar”), subtle image artifacts emerge when using conventional $B_0$-mapping-based calibration, whereas the proposed grouped-kernel calibration continue to yield clean reconstructions. In the coronal plane, significant artifacts appear in reconstructions using the $B_0$-mapping-based calibration, likely caused by eddy currents induced on components like the RF coil shield, particularly along the z-axis. In contrast, the proposed auto-calibration method produces consistently clean reconstructed images.

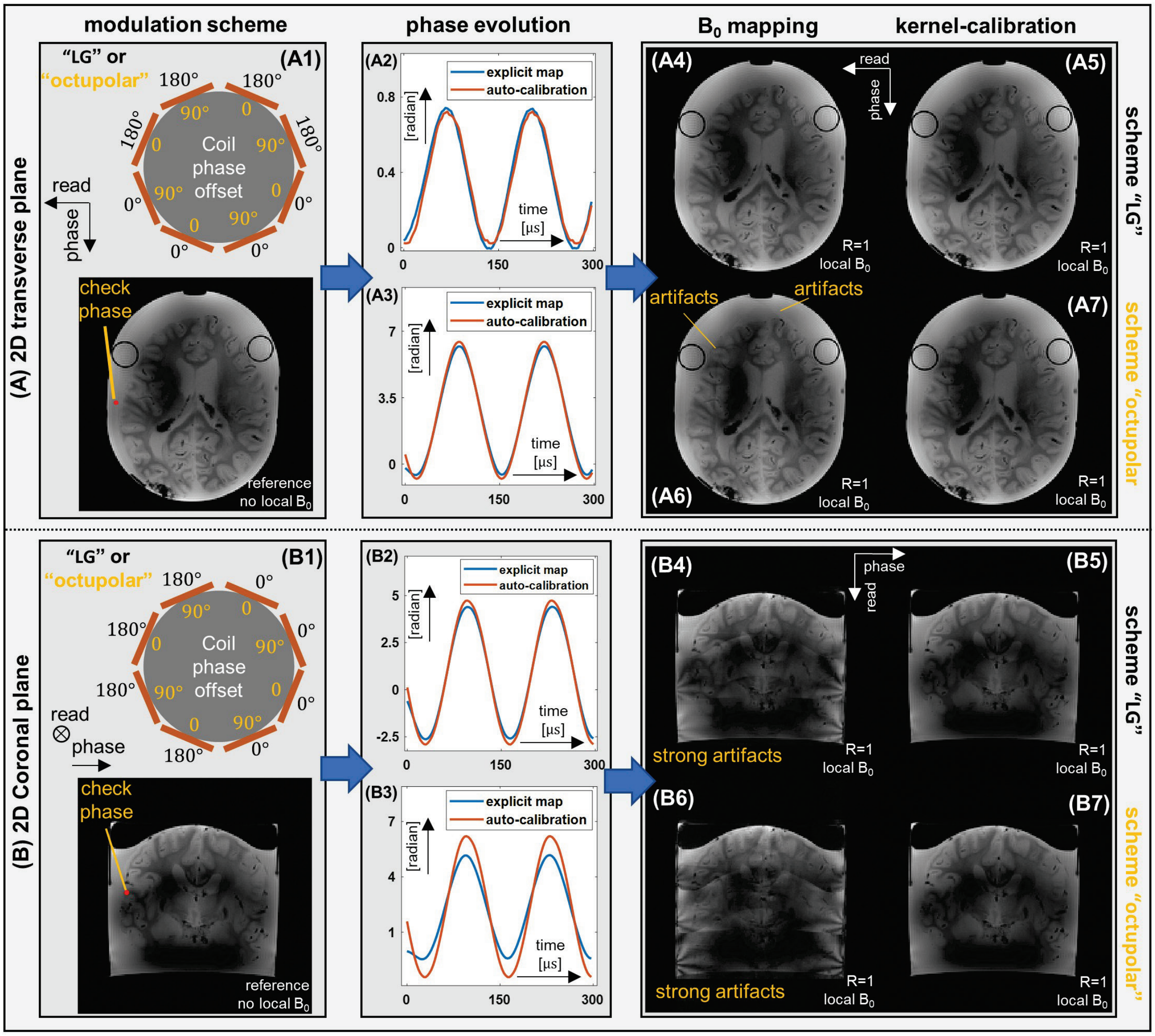

Figure 5. Ex-vivo 2D FLASH with rapid modulation of local $B_0$ coil array, comparing the proposed auto-calibration method, and a previous technique that obtains phase evolution maps with $B_0$ mapping, current monitoring and ESPIRiT high-resolution grid

extrapolation. (A) Transverse plane. (B) Sagittal plane. Two modulation schemes “LG” (nearly linear gradient) and “octupolar” (higher-order nonlinear gradient) are shown, with different readout orientations. (A1, B1) the phase offsets of sinusoidal currents in local $B_0$ coils, and the reference image without $B_0$ modulations. (A2-A3, B2-B3) The phase evolution of the encoding matrix due to local $B_0$ modulations at a selected “check-phase” pixel, obtained from the two calibration methods. Linear phase ramp due to DC offsets from coil currents and mean phase differences are removed to visually compare the oscillation of the phase curves only, that is mostly relevant to residue folder-over artifacts. (A4-A7, B4-B7) The reconstruction of fully-sampled k-space using two field-calibration techniques. The $B_0$ mapping based calibration can lead to artifacts especially along the z axis. The grouped-kernel auto-calibration achieved robust reconstruction across different spatial axes, and complex nonlinear modulation fields.

### 4.2 Temporally or spatially decoupled local gradients

In Figure 6-7, one important application of grouped-kernel calibration is to enable robust experimental comparisons across distinct field modulations for image acceleration. Without our auto-calibrations, such comparisons can be vulnerable to possible artifacts, as in (A6, B4, B6) of Figure 5. In previous study[24], the 8-channel local $B_0$ coils optimized for 2D Cartesian transverse scans generated a nearly-linear gradient oscillation along the phase-encoding dimension, converging nonlinear gradient modulations to Bunched-phase-encoding. This configuration was proved more efficient than others for the given coil geometry, likely because it concentrated all the “sampling energy” of the modulation functions into the only one undersampled dimension with the induced zig-zag trajectory, thereby avoiding unnecessary modulation along the fully-sampled readout direction.

However, maintaining the global gradient linearity may not be as critical as keeping the proper gradient modulation direction, which thus, provides flexible degrees of freedom in local $B_0$ encoding. Therefore, leveraging on auto-calibration, the optimized field modulation schemes in transverse and coronal FLASH scans was further investigated, comparing a conventional globally-linear gradient with two local gradients that are either temporally or spatially decoupled. In Figure 6, the two sets of $B_0$ coils positioned at the opposite sides along phase-encoding axis, can together form a globally-linear gradient when driven by sinusoidal currents with 180-deg phase shifts. However, when temporally alternating their activations with 90-deg phase shifts to break the global spatial-linearity, the resulted local gradient modulation scheme achieved very similar least-square reconstruction quality and minimal difference in mean values of G-map and power function maps. Similarly, in Figure 7, the relative z-position between the ex-vivo phantom and the local $B_0$ coil array was shifted, to produce a globally linear gradient or spatially-swapped local gradients with different polarity but similar gradient strengths preserved in local regions (e.g., the four corners of the FOV). Both global and local linear gradient schemes provided similar modulations along phase-encoding dimension within local regions, and can reach similar sampling efficiency and reconstruction quality. Both examples illustrate that, strict global linearity may not be the only option for achieving effective image accelerations, as experimentally verified by our grouped-kernel calibration technique.

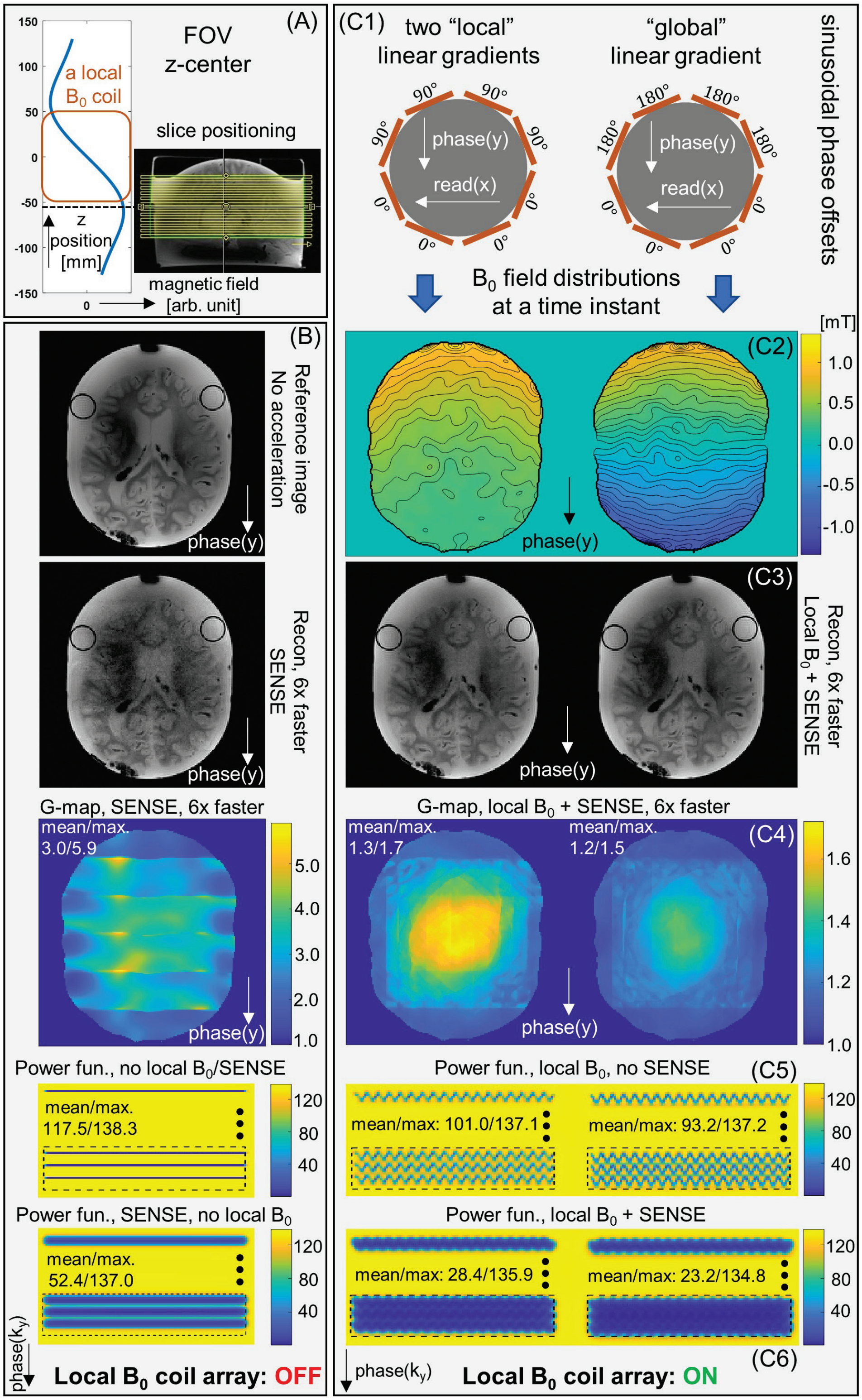

Figure 6. Ex-vivo scan in the 2D transverse plane, comparing temporally-decoupled local linear gradients and a global linear gradient. (A) a slice near the maximum modulation field is shown. (B) The reference image with fully-sampled k-space, the SENSE-only 6x accelerated image and G-map are shown. In k-space, the power function (approximation errors) maps with 6x

undersampling with and without multiple RF receivers are shown. (C1) The phase offsets of sinusoidal currents in local $B_0$ coils are shown, for synthesizing the local and the global linear gradients along the phase-encoding dimension, respectively. (C2) The magnetic field distributions with iso-contour (i.e., indicating gradient directions) at a time instant (nearly current peak) are shown, for the two modulation schemes. (C3) The reconstruction of 6x joint acceleration of local gradients and SENSE is only slightly noisier than the one using the global gradient, while both are substantially improved than the SENSE-only accelerated image. (C4) The G-map by the local gradients is only slightly worse than the global gradient, while both are substantially reduced than the SENSE-only one. (C5) The k-space power function maps for the two modulation schemes without SENSE are shown, where the global gradient induces a slightly darker zig-zag trajectory corresponding to subtly reduced approximation errors. (C6) The power function maps for the two modulation schemes combining SENSE, where the difference in the effective k-space coverage is little. For both schemes, hot-spots in SENSE-only acceleration are removed by the dynamic $B_0$ kernels.

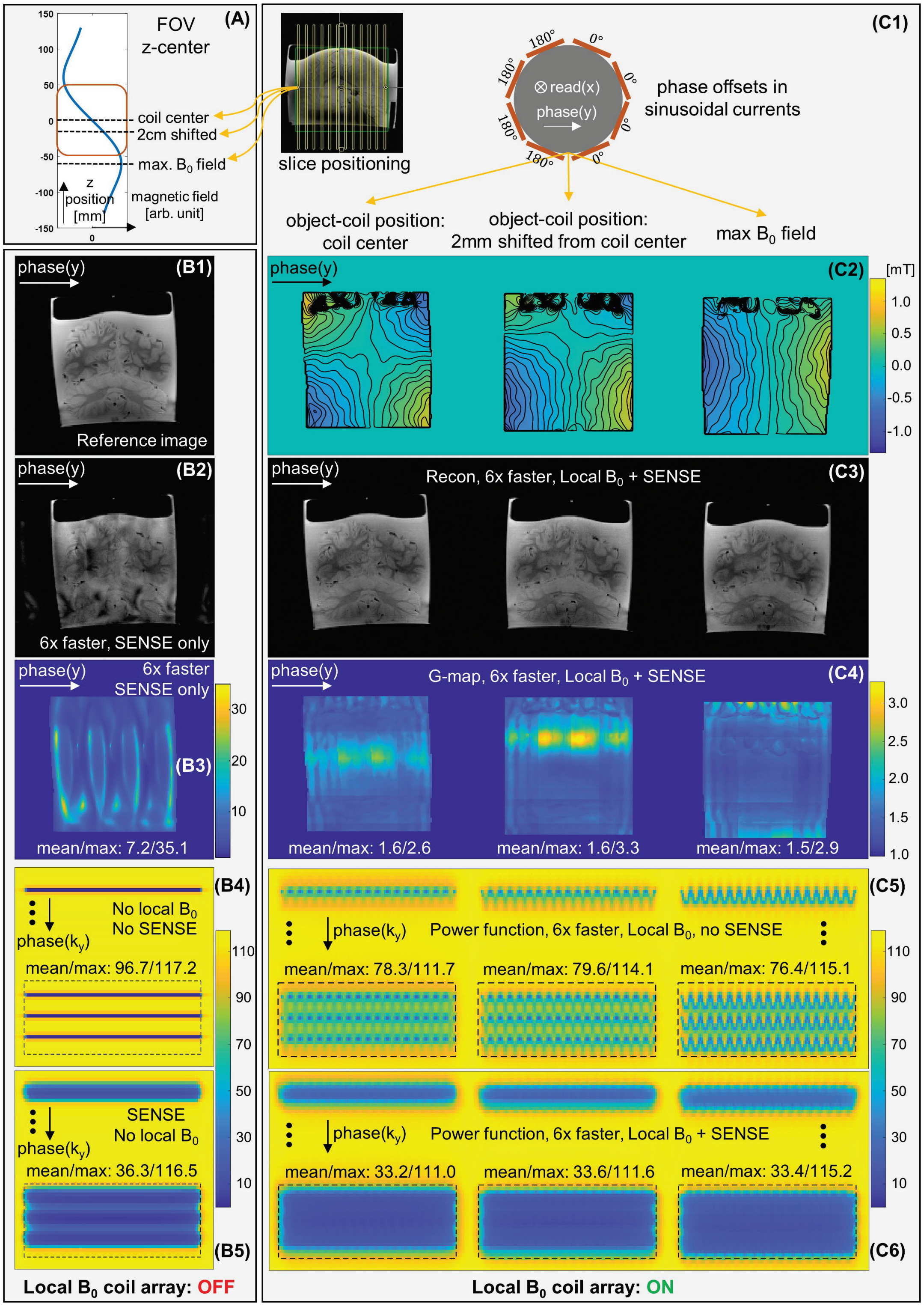

Figure 7. Ex-vivo scan in the 2D coronal plane, comparing sampling efficiency when shifting between spatially-decoupled local gradients and a global linear gradient. (A) The FOV centers along z for generating different $B_0$ field shapes are shown. (B1-B5) The reference image with fully-sampled k-space, the SENSE-only 6x accelerated image and G-map are shown. In k-space, the power function (approximation errors) maps with 6x undersampling with and without multiple RF receivers are shown. (C1) The phase offsets of sinusoidal currents in local $B_0$ coils are shown, for synthesizing the local and the global linear gradients along the phase-encoding dimension, respectively. (C2) The magnetic field distributions with iso-contour (i.e., indicating gradient directions) at a time instant (nearly current peak) are shown, for the three modulation schemes. (C3) The reconstruction of 6x joint acceleration of local gradients and SENSE are similar to the one using the global gradient, while all of them are substantially improved than the SENSE-only accelerated image. (C4) The G-maps by the local gradients are close to the global gradient, while both are substantially reduced than the SENSE-only one. (C5) The k-space power function maps for the three modulation schemes without SENSE are shown, where the global gradient induces a zig-zag trajectory, and when moving to the leftmost case (shifting along z-position), induces sampling using a similarly efficient "stamp". (C6) The power function maps for the two modulation schemes combining SENSE, where the difference in the effective k-space coverage is little. For all schemes, hot-spots in SENSE-only acceleration are removed by the dynamic $B_0$ kernels.

### 4.3 Robust field calibration against hardware spike noise

Figure 8 illustrates the robustness of our proposed grouped-kernel field calibration against hardware noise interference and subtle motions during in-vivo scans. In this case, the field-modulated data were affected by occasional spike noise (visible in RF chains and raw data) caused by malfunctioning local $B_0$ coils. Additionally, there was slight rotation between the standard and field-modulated scans performed with full k-sampling (each 2mins, >6mins gap between).

Although the field-modulated ACS data were consequently corrupted, the proposed kernel-calibration technique remained accurate in both fully-sampled and 4x3-fold undersampled reconstruction. The solution of each kernel was constrained by a large number of interpolation pairs between the standard and field-modulated ACS regions, across numerous readouts, phase-encoded steps and RF receiver channels, far exceeding the number of unknowns to be resolved per kernel. This kernel-encoded redundancy in k-space data enabled crucial averaging effects and suppressed the influence of such localized spike noise for field calibrations, inheriting the calibration robustness of RF sensitivity kernels/maps in GRAPPA/ESPIRiT.

In contrast, this important noise-resistance feature is absent in the other full-model field calibration techniques, such as the CSI-based calibration scan, and the image-space auto-calibration which estimates the PSF spreading relationships between reference and field-modulated high-resolution images on a pixel-by-pixel basis.

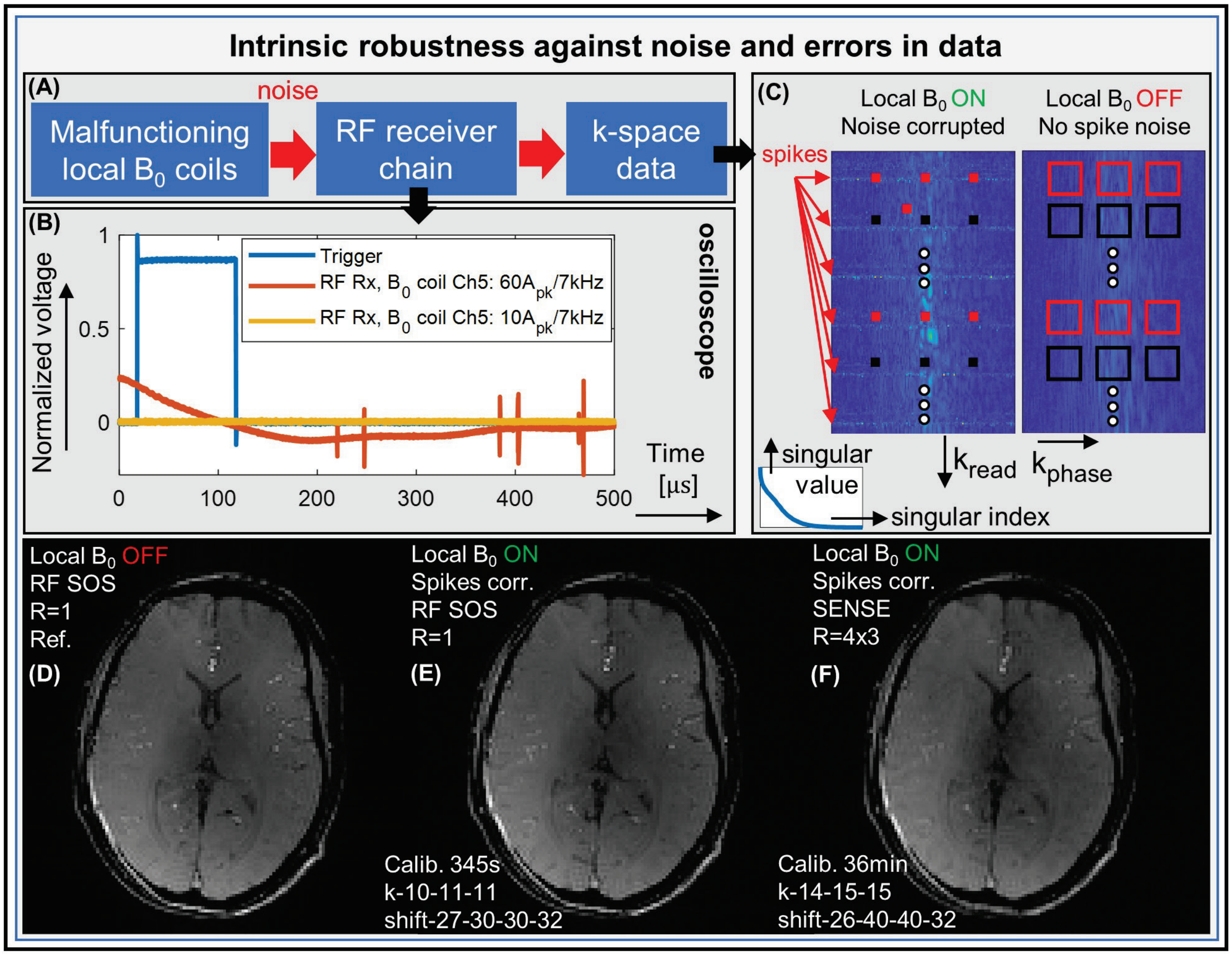

Figure 8. Noise resilience. (A) Malfunctioning local $B_0$ coils injected additional spike noise into the RF receiver chain. (B) The noise was detected by a high-speed oscilloscope probing the RF receiver chain (after pre-amp), when the sinusoidal coil current increases from $10A_{pk}$ (useless for modulation experiments) to $60A_{pk}$, possibly caused by e.g., electrical sparks. (C) Noise bursts are visible in the raw k-space data (field-modulated ACS), interfering the kernel estimation. (D, E, F) Despite this, both the fully-sampled and 4x3-fold accelerated reconstructions (joint $B_0$+SENSE) remain clean without apparent residual artifacts. Subtle inter-scan (between reference and field-modulated) head rotations are also present, which further validates calibration stability.

### 4.4 In-vivo image acceleration

Figure 9 provides a comprehensive summary of in-vivo 2D FLASH (coronal plane, phase-encoding: head-feet axis) with optimized local $B_0$ modulation scheme (0-deg), to evaluate in-vivo reconstruction quality and sampling efficiency. Driven by in-phase sinusoidal currents, the local $B_0$ coils provided a nearly-linear z-gradient along phase-encoding dimension (note, different than Section 4.2), approximately realizing 2D bunched-phase-encoding. Joint acceleration using rapid $B_0$ modulation and multiple RF receivers (SENSE) yielded clean $L_2$ reconstruction from 7x undersampled k-space without noticeable residue artifacts but only subtle noise amplification, clearly outperforming SENSE-only acceleration. The G-factor map shows a mean/max of 1.39/2.88 and decreases further towards peripheral slices. Power (approximation error) and cardinal (noise amplifications) functions maps indicates smooth, well-covered k-space without abrupt hot spots. Additional compressed-sensing $L_1$ reconstruction further optimized the reconstruction quality, which was computed efficiently with joint data compression based on both $B_0$ and $B_1$ kernels (approx. total 10s/slice).

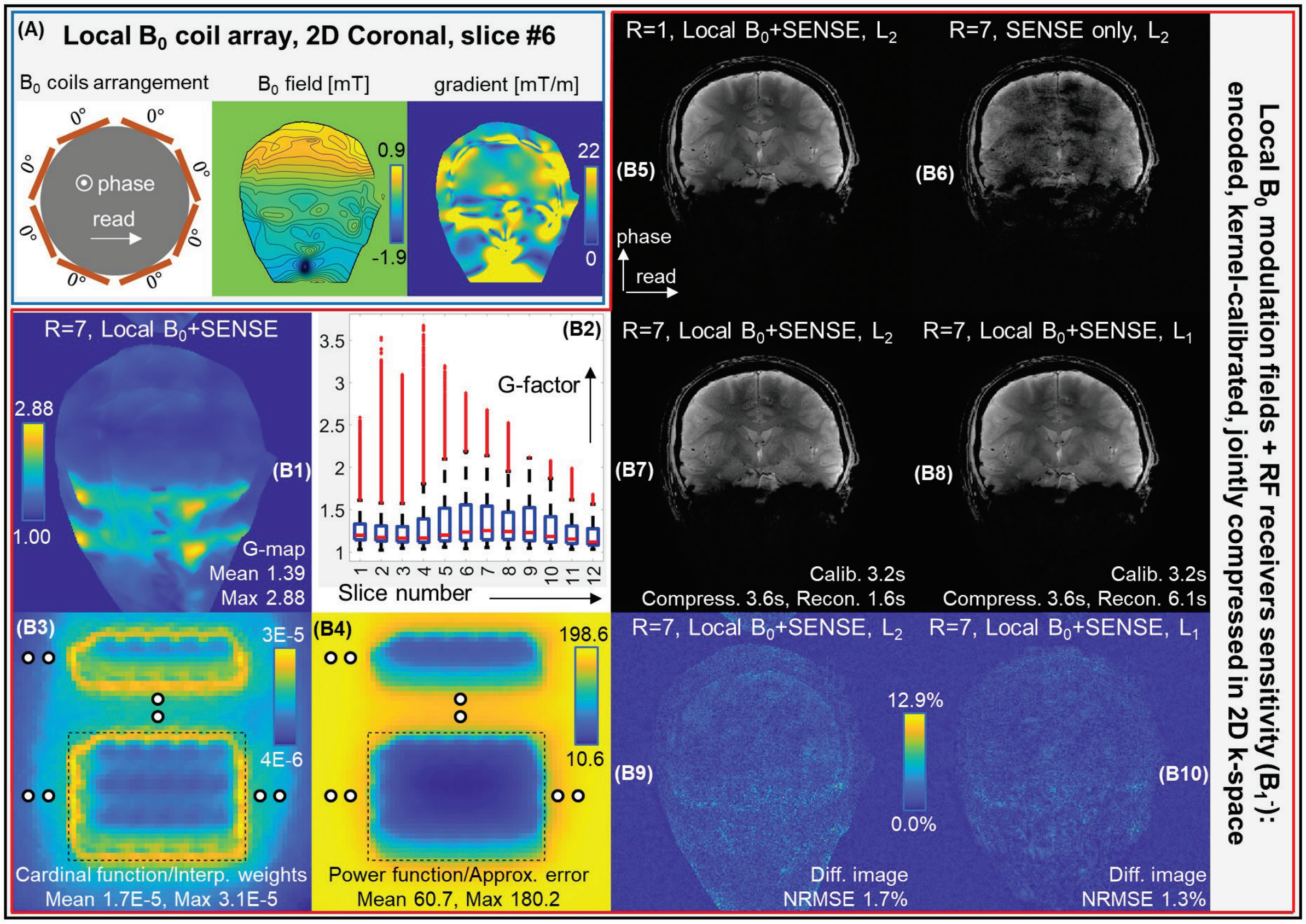

Figure 9. (A): The phase offsets in local $B_0$ coils, the magnetic field distribution generated at nearly-peak current, and the estimated gradient strength distribution (Euclidean-norm in two orthogonal dimensions) are shown. (B1-B4): The G-map, the G-factor boxplot for each of 12 slices, the k-space cardinal and power function maps are shown, which indicate sufficient sampling to reach 7x joint acceleration factor by $B_0$ modulations and SENSE. (B5-B10): The reference fully-sampled image with B0 and SENSE reconstruction is successfully reconstructed. The $L_2$ reconstruction of 7x undersampled k-space has only subtle noise amplifications without noticeable residue fold-over artifacts. Its $L_1$ reconstruction further improves the image quality, as also shown in their difference images, and is very close to the reference image. In comparison, the SENSE-only acceleration has severe artifacts.

Similarly, Figure 10 provides a comprehensive summary of in-vivo 3D FLASH with optimized local $B_0$ coil array and 3D Wave-CAIPI. The 4x3-fold joint acceleration of local $B_0$ coils and SENSE produced clean $L_2$ reconstruction images, with minimal fold-over residue artifacts, and only subtle noise amplification. The $B_0$ modulation field shape was approximately an off-center quadratic function in the transverse plane, and nearly-linear gradient along z-axis. The compressed-sensing $L_1$ reconstruction further improved the image quality, and was computed efficiently with joint data compression based on both $B_0$ and $B_1$ kernels (approx. 7 mins total/volume).

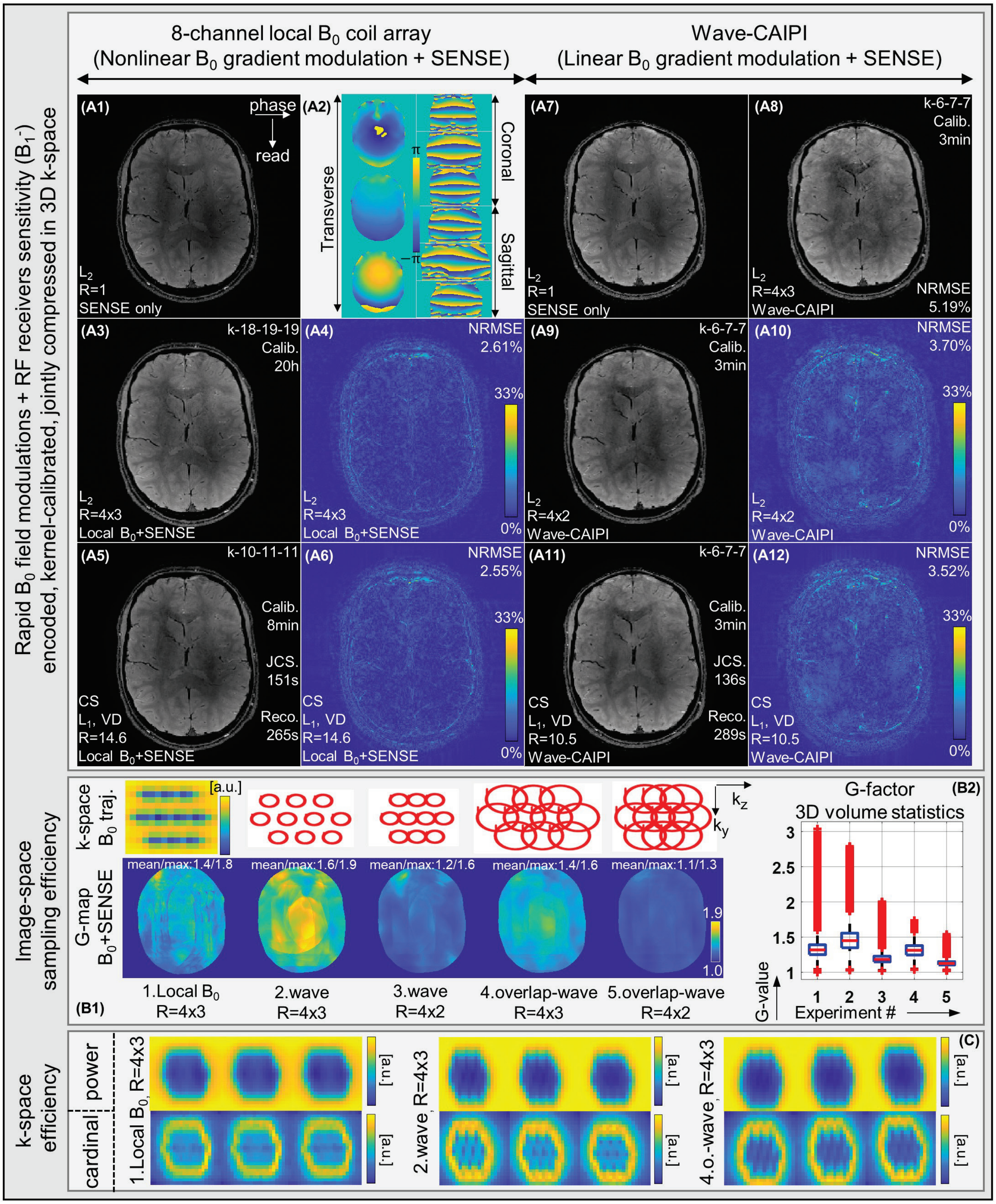

Figure 10. (A) 3D in-vivo MRI accelerated by local $B_0$ coils (nonlinear gradients) and Wave-CAIPI (linear gradients), together with SENSE. (A1, A7): The reference fully-sampled SENSE images without $B_0$ modulations. (A2): The additional spin phase evolution maps (near-peak) caused by local $B_0$ coil array are shown for transverse, coronal and sagittal planes, at the third-dimension locations ¼ FOV, ½ FOV, -¼ FOV. They are transverse off-center quadratic functions combined with z-gradients. (A3-A6): The 4x3-fold joint accelerated FLASH has only minor noise amplifications in $L_2$ reconstructions, with further optimized image quality in $L_1$ reconstruction with 14.6-fold variable-density undersampling. The difference images are mostly confined to vessels (low signal), edges and fat off-resonance regions. (A8): The 4x3-fold joint accelerated Wave-CAIPI with non-overlapped wave trajectory, with residue fold-over artifacts in $L_2$ reconstruction. (A9-A12): The 4x2-fold joint accelerated Wave-CAIPI has minor noise amplifications in $L_2$ reconstructions, and the image quality is further improved in the $L_1$ reconstruction with 10.5-fold variable-density undersampling. The difference images have errors in the low-frequency regions additional to edges and fat off-resonance areas, possibly due to subject motions. (B1): The k-space $B_0$ trajectories for five 3D in-vivo experiments are shown, including the k-space "stamp" by local $B_0$ array, and the pointwise wave-trajectories displayed in Pulseq, under different undersampling factors. The slice-specific G-maps are shown that, the nonlinear gradients acceleration efficiency is nearly identical to the large "Wave" modulations (overlap-wave, R=4x3). (B2) The boxplot for the whole-volume G-factor (excluding 10 oversampled edge slices) is shown, where nonlinear gradient encoding (Experiment #1) reaches similar efficiency ($25^{th}$ – $75^{th}$ percentiles) as the overlapped-wave trajectory (Experiment #4) given 4x3-fold acceleration. (C): The k-space cardinal and power function maps for 4x3-fold acceleration using nonlinear, non-overlapped wave, and overlapped-wave modulations, without displaying in a common scale. The nonlinear gradient modulations produce a smoother and slightly broader k-space coverage; however, an upward scaling is found, which might due to the summed interpolation weights (Frobenius norm) additionally over readout dimensions (3D kernels). The Wave-CAIPI trajectories show more non-uniform k-space coverage. More detailed interpretations are reserved for future investigations.

For comparison, the 4x2-fold accelerated wave-CAIPI with a non-overlapped wave trajectory also yielded clean $L_2$ and $L_1$ reconstructions. Although the overlapped wave trajectory with larger gradient currents was nominally capable of 4x3-fold acceleration, the resulting reconstructions were slightly degraded, likely related to strong inter-scan variations (>13mins gap between), and was mainly used for G-map comparisons.

In Figure 10(B1), the local $B_0$ modulation required larger calibration kernels to achieve artifact-free reconstruction than Wave-CAIPI, which uses a conventional pointwise k-space ($B_0$) trajectory. Notably, the volume G-factor ($25^{th}$-$75^{th}$ percentiles) for 4x3-fold acceleration with the nonlinear $B_0$ gradients was nearly identical as the same acceleration factor achieved with scanner's modulation gradients (both non-overlapped and overlapped waves). This again suggests that global linearity of the modulation gradients is not always required to achieve comparable acceleration performance.

In Figure 10(C), the power (approximation error) and cardinal (noise amplification) function maps show a smoother, slightly broader k-space coverage by local $B_0$ coils plus SENSE, than Wave-CAIPI (with SENSE). However, these maps are shown on different scales. The maps for local $B_0$ coils appear scaled upward, possibly reflecting additional non-zero interpolation weights from the readout-dimension, which requires further investigations in the future.

### 4.5 Group-patch RKHS reconstruction: k-space subregion-wise interpolation

In Figure 11, the group-patch reconstruction was implemented and compared with the hybrid-space reconstruction (as in conventional Wave-CAIPI), as proof-of-concept without joint SENSE acceleration, data compression and parallelization.

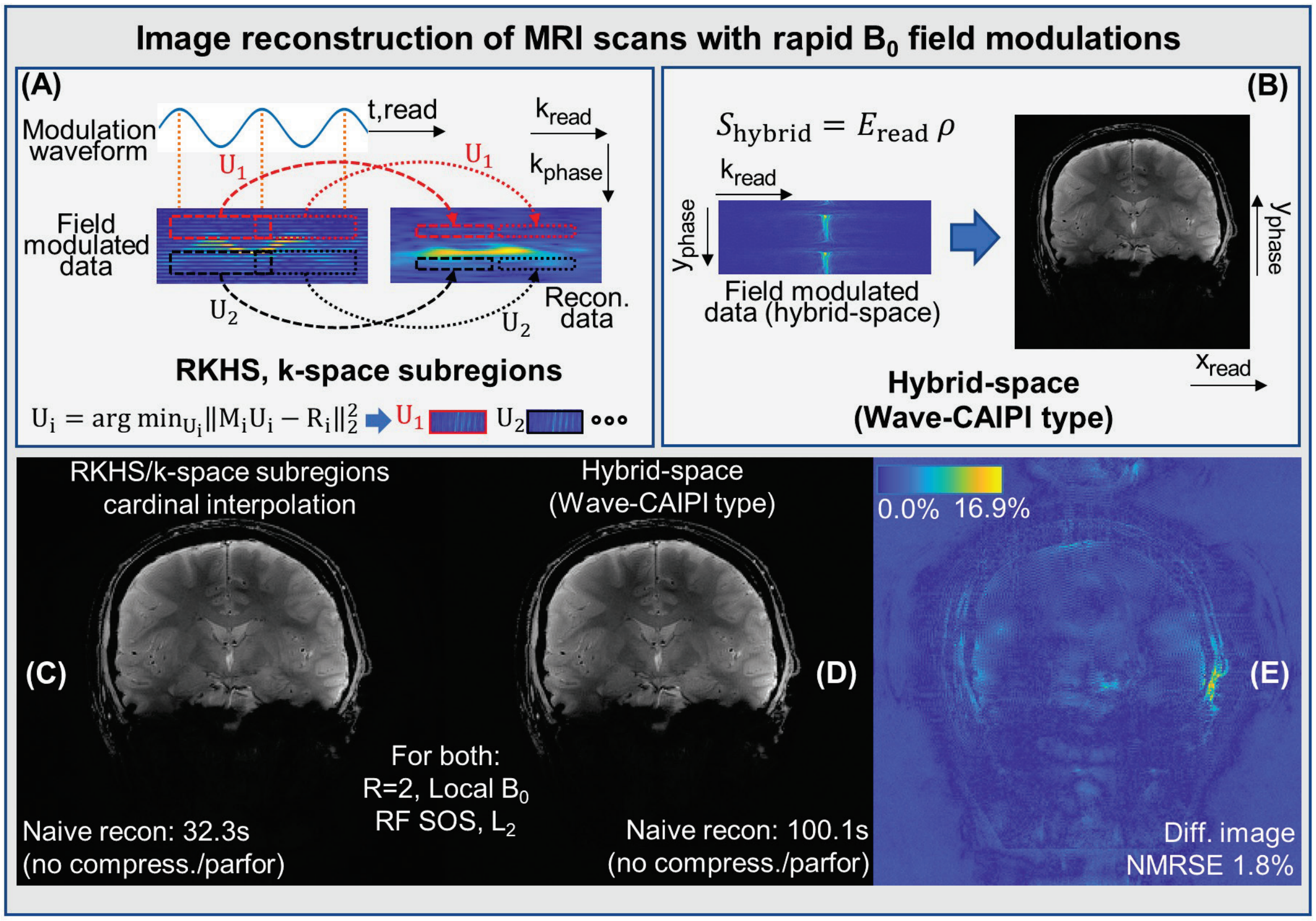

Figure 11. Comparison: recover the k-space data through RKHS interpolations and then yield the image through FFT, or reconstruct the image through hybrid-space as in Wave-CAIPI. (A) In group-patch reconstruction, the k-space data are decoupled in subregions, spanning periodically occurring $B_0$ kernels, along both phase-encoding and readout dimensions. Cardinal functions for distinct subregions (index i) are calculated based on the RKHS formalism, and applied for k-space interpolation. Appropriately defined subregion can further reduce peak memory and reconstruction time. (B) In hybrid-space reconstruction, the phase-encoding dimension(s) are solved by FFT, while the readout forward model is solved by general

inversion. Note, the k-space dataset and the reconstructed image only serves as illustration purposes, and size-ratio between their figures is not realistic. (C-D) The group-patch and hybrid-space reconstruction images are shown. Such cardinal function interpolation based on k-space subregions achieved about 3x lower computational time than solving forward models in hybrid-space, given 2x undersampled k-space, without compression, parallel computing and SENSE joint accelerations. (E) The difference image shows a subtle fold-over pattern. However, more detailed comparisons after combining joint $B_0$-$B_1$-kernel compression are reserved for future work. Note, the kernel calibration time, joint compression time, and reconstruction time are shown.

This proposed k-space interpolation-based reconstruction achieved 3x less computational time than solving the forward model in hybrid-space, by interpolating the acquired 2x undersampled k-space onto a readout-oversampled Fourier grid, in multiple subregions spanning several phase-encoded steps and a readout modulation period. A total of 24 cardinal function matrices were computed and reused, while several others were computed once for some k-space edges.

Theoretically, such k-space group-patch reconstruction can further support joint acceleration by rapid $B_0$ fields and RF receivers modulations, with parallel computing and joint compression of undersampled k-space data, which is reserved for future work. The joint compression technique will be described in a separate paper, although it has been applied in results of Figure 9 and 10 indicated by “JCS.”.

**Discussions and conclusions**

According to the ex-vivo and in-vivo scans with rapid $B_0$ field modulations, this paper demonstrates that, the grouped-kernel continuous field calibration technique is robust to noise and various system imperfections. Additionally, the RKHS-based, group-patch reconstruction is feasible and efficient.

Such data-driven calibration enabled experimental tests of various $B_0$ modulation schemes along different FOV axes, which otherwise may become unrealistic e.g., by suffering from complex eddy currents. Global spatial linearity of the modulation gradients is not always required to achieve comparable image acceleration, complementing conclusions in the previous study. Although using our 8 local $B_0$ coils, nonlinear modulation gradients that can substantially surpass linear ones have not been found yet, these findings can motivate further hardware designs that can generate more localized $B_0$ modulation fields.

A 3D nonlinear field ($35_{pk}$ ampere per coil) as a combination of transverse quadratic field[45] and z-linear gradient delivered similar acceleration performance as Wave-CAIPI. The latter uses scanner gradient modulations, which relies on more elaborate wire patterns and higher electric power to impose linear gradients on the region of interest from the far away magnet bore. While both nonlinear and linear gradients might not be exhaustively optimized (e.g., modulation amplitudes, frequency), the grouped-kernel calibration has shown to be universally-applicable, and may support further searches over different gradient designs (e.g., more localized nonlinear gradients, high-performance head gradients) and different readout trajectories (high-bandwidth GRE, EPI[46]).

All in-vivo data were retrospectively undersampled and cropped for ACS region, so subject motion was inevitable and likely larger than in real-time accelerated scans with integrated ACS. Even under these conditions, the grouped-kernel calibration remained stable, robustly extracting the spatiotemporal $B_0$ modulations via redundant k-space interpolations. Additionally, receiver-combined ACS data may be explored for faster kernel estimation. As matrix size increases, the conventional 3D Wave-CAIPI reconstruction can be substantially hindered with respect to speed and memory. However, interpolating k-space data stream in small 3D chucks (subregions) using a small reusable set of cardinal functions appears as a promising solution.

Notably, rather than advocating a specific sequence (e.g., BPE, Wave-CAIPI, a particular nonlinear field), the emphasis of this paper is on elucidating the principle of kernel calibration and reconstruction in k-space groups or subregions. To explicitly extract kernels, interpolation relationships can be established anywhere in k-space and grouped by instantaneous image-space modulation, as long as the two carefully-selected ACS regions differ only in the spatial modulations to estimate. With fully-sampled kernels or spin spatial-encoding maps, multidimensional data can be interpolated onto a reconstruction Fourier grid without being tied to the exact form of the learned local kernel.

Beyond rapid modulations (e.g., min. a few kHz) of nonlinear (e.g., local $B_0$ coils, FRONSAC) and linear $B_0$ gradients (BPE, Wave-CAIPI), the principle underlying the grouped-kernel calibration or reconstruction techniques might be adapted for other methods with critical necessity to explicitly extract spatial modulation functions (e.g., EPI, radial, fm-bSSFP, RF encoding, shot-to-shot signal variations[47], subpixel encoding[48,49]). In future study, while our approaches and other kernel-based MRI methods may be extended to broader scenarios, their connections to other statistical learning algorithms backboned by the RKHS theory is also interesting to explore.

Essentially, inspired by the RKHS view, nonlinear and linear $B_0$ modulations can now be encoded[24], kernel-auto-calibrated[34], jointly compressed, and reconstructed entirely in k-space, unifying the mathematical foundations of GRAPPA ($B_1$ kernels) and nonlinear gradients ($B_0$ kernels) encoding. Namely, the oversampled readout induces an additional acceleration dimension that is mathematically-equivalent to parallel imaging acceleration, pushing towards the ultimate MRI speed in a practical and reliable manner.

## Appendix - Algorithm implementation examples

### A1 Grouped-kernel continuous field calibration

Step 1: Acquire ACS pairs. Obtain two low-resolution fully-sampled ACS datasets, e.g., a standard ACS and a field-modulated ACS, either by integrated acquisition or by retrospective cropping of fully-sampled scans.

Step 2: Roughly estimate trajectory and kernel size. Make an initial guess of the field-modulated trajectory (e.g., peak amplitudes of the drifted $B_0$ kernels) to guide grouping of k-space points by their instantaneous spatial-modulations.

Step 3: Construct interpolation systems in groups. For each spatial-modulation state (group) at all readout and phase-encoding k-space locations, and all RF receiver channels, slide a mathematical window to collect source neighborhoods in the standard ACS and target points in the field-modulated ACS, e.g., to form a system of linear equations. Reverse direction interpolation may also be applicable.

Step 4: Solve each group's interpolation relationships with a numerical solver, e.g., pseudo-inverse, tSVD, LSQR, or neural networks, to obtain a shift-invariant kernel for that spatial-modulation state.

Step 5: repeat for all groups separately – possible with parallel computing – to recover a complete set of dynamic $B_0$ kernels. If full Fourier-grid kernels are required, they can be zero-filled and Fourier transformed to derive spatial phase evolution maps.

Step 6: The estimated time-varying kernels or a full spatiotemporal model of phase evolution maps can be used for image reconstruction, either through hybrid-space as in Wave-CAIPI, or via RKHS-based k-space interpolation. Theoretically, if interpolating kernels from field-modulated ACS to standard ACS, the SPIRiT-type reconstruction is also possible.

**A2 Group-patch reconstruction with subregion-wise k-space interpolation**

Step 1: Partition undersampled k-space data and reconstruction Fourier grid. Divide the full k-space grid into subregions, each span one or a few periodically occurring $B_0$ kernels, maximizing the reuse of interpolation weights later.

Step 2: Compute cardinal functions. For each subregion, formulate the RKHS interpolation equations as Equation (1-3) using the spatial-encoding maps at the source-subregion's k-space timepoints, and the reference spatial-encoding maps (e.g., DFT matrix) at the target-subregion's k-space timepoints. Solve a cardinal function matrix that maps a slightly larger (with margins) "source" region in undersampled k-space data, onto a target subregion on reconstructed Fourier grid. These cardinal function matrices can be reused for any other subregions sampled by the same set of $B_0$ kernels, sharing identical interpolation relationships between the source and target regions.

Step 3: Apply the computed cardinal function matrices to map each source subregion in undersampled data onto its target subregion in the reconstruction Fourier grid.

Step 4: Obtain the reconstructed image by Fourier transforming the reconstructed k-space data.

**Data availability**

The codes and example data will be available in public once the paper is accepted.

## Acknowledgement

This study is supported by the ERC Advanced Grant (No. 834940).

The ex vivo brain phantom was with courtesy of the Institute of Clinical Anatomy and Cell Analysis, Department of Anatomy, Eberhard Karls University of Tübingen. The first author thanks Dr. Thomas Shiozawa (Institute of Clinical Anatomy and Cell Analysis) for assistance with sample preparation, and Dr. Gisela Hagberg for assistance in scanning this phantom.

The first author also thanks Stefan Plappert for guidance in programming the ADwin high-speed processor, Dr. Felix Breuer and Prof. Berkin Bilgic for helpful discussions about CAIPIRINHA reconstructions, Dr. Georgiy Solomakha, Dr. Nikolai Avdievitch, Dario Bosch and Dr. Yi Chen for assistance in debugging the hardware noise injected into the RF receiver system, Dr. Anagha Deshmane for discussions about GROG kernels.

We are grateful to the late Mirsat Memaj (Fine mechanical workshop, Max Planck Institute for Biological Cybernetics) for manufacturing the coil support of the 8-channel local $B_0$ coil array.